\documentclass[useAMS, usenatbib]{mn2e}
\usepackage{lscape}
\usepackage{rotating}
\usepackage{amssymb}
\usepackage{float}
\usepackage{hyperref}
\hypersetup{colorlinks = true, linkcolor = blue, urlcolor=blue, citecolor=blue} 
\usepackage[normalem]{ulem}
\newcommand{\GG}[1]{}

\def\msun{\hbox{M$_\odot$}}

\def\frac{\hbox{f$_{\rm mix}$}}

\def\t4{\hbox{t$_{\rm 4}$}}

\def\cm3{\hbox{cm$^{-3}$}}

\input psfig.sty
\input epsf.sty
\usepackage{graphicx}
\usepackage{epsfig}
\def\cubi{$C_{F336W, F438W, F814W}$}
\def\cunbi{$C_{F343N, F438W, F814W}$}
\def\cubifancy{\texttt{CUBI}}
\def\cunbifancy{\texttt{CUnBI}}
\def\wrgb{$W_{\rm RGB}$}
\def\wrc{$W_{\rm RC}$}
\def\wagb{$W_{\rm URGB}$}

\title[No Evidence for MPs in NGC 419]
{The Search for Multiple Populations in Magellanic Cloud Clusters III: No evidence for Multiple Populations in the SMC cluster NGC 419}
\author[Martocchia et al.] {S. Martocchia$^{1}$, N. Bastian$^{1}$, C. Usher$^{1}$, V. Kozhurina-Platais$^{2}$, F. Niederhofer$^{2,3}$,
\newauthor I. Cabrera-Ziri$^{1}$, E. Dalessandro$^{4}$,  K. Hollyhead$^{1}$, N. Kacharov$^{5}$, C. Lardo$^{1}$,   
\newauthor S. Larsen$^{6}$, A. Mucciarelli$^{7}$, I. Platais$^{8}$, M. Salaris$^{1}$, M. Cordero$^{9}$, D. Geisler$^{10}$,
\newauthor M. Hilker$^{11}$, C. Li$^{12}$, D. Mackey$^{13}$\\
$^{1}$Astrophysics Research Institute, Liverpool John Moores University, 146 Brownlow Hill, Liverpool L3 5RF, UK\\
$^{2}$Space Telescope Science Institute, 3700 San Martin Drive, Baltimore, MD 21218, USA\\
$^{3}$Leibniz-Institut f\"ur Astrophysik Potsdam, An der Sternwarte 16, Potsdam 14482, Germany\\
$^{4}$INAF, Osservatorio Astronomico di Bologna, via Ranzani 1, 40127, Bologna, Italy\\
$^{5}$Max-Planck-Institut f\"ur Astronomie, K\"onigstuhl 17, D-69117 Heidelberg, Germany\\
$^{6}$Department of Astrophysics/IMAPP, Radboud University, P.O. Box 9010, 6500 GL Nijmegen, The Netherlands\\
$^{7}$Department of Physics and Astronomy, University of Bologna, Viale Berti Pichat 6/2, I-40127 Bologna, Italy\\
$^{8}$Department of Physics and Astronomy, Johns Hopkins University, 3400 North Charles Street, Baltimore, MD 21218, USA\\
$^{9}$Astronomisches Rechen-Institut, Zentrum f\"ur Astronomie der Universit\"at Heidelberg, M\"onchhofstrasse 12-14, D-69120 Heidelberg, Germany\\
$^{10}$Departamento de Astronomia, Universidad de Concepcion, Casilla 160-C, Chile\\
$^{11}$European Southern Observatory, Karl-Schwarzschild-Stra\ss e 2, D-85748 Garching bei M\"unchen, Germany\\
$^{12}$Department of Physics and Astronomy, Macquarie University, Sydney, NSW 2109, Australia\\
$^{13}$Research School of Astronomy and Astrophysics, Australian National University, Canberra, ACT 2611, Australia\\
}
\date{Accepted. Received ; in original form.}
\pagerange{\pageref{firstpage}--\pageref{lastpage}}
\pubyear{2017}

\begin{document}

\maketitle
\label{firstpage}

\begin{abstract}	
We present the third paper about our ongoing HST survey for the search for multiple stellar populations (MPs) within Magellanic Cloud clusters. We report here the analysis of NGC 419, a $\sim 1.5$ Gyr old, massive ($\gtrsim 2 \times 10^5$ \msun) star cluster in the Small Magellanic Cloud (SMC). By comparing our photometric data with stellar isochrones, we set a limit on [N/Fe] enhancement of $\lesssim$+0.5 dex and hence
we find that no MPs are detected in this cluster. 
This is surprising because, in the first two papers of this series, we found evidence for MPs in 4 other SMC clusters (NGC 121; Lindsay 1, NGC 339, NGC 416), aged from 6 Gyr up to $\sim 10-11$ Gyr. This finding raises the question whether age could play a major role in the MPs phenomenon. 
Additionally, our results appear to exclude mass or environment as the only key factors regulating the existence of a chemical enrichment, since all clusters studied so far in this survey are equally massive ($\sim 1-2 \times 10^5$ \msun) and no particular patterns are found when looking at their spatial distribution in the SMC. 
\end{abstract}


\section{Introduction}
\label{sec:intro}
In the last few decades, our understanding of globular clusters (GCs) as simple stellar populations (SSPs) has been overturned by the presence of star-to-star abundance spreads in light elements (e.g., C, N, O, Na) and by the broadening/splitting of features in clusters' colour-magnitude diagrams (CMDs). 

The presence of multiple stellar populations (MPs) in GCs appears to be nearly ubiquitous in the most nearby galaxies. Large numbers of Milky Way GCs have been studied \citep{gratton12} and all old ($>10$ Gyr) clusters surveyed so far have been found to host MPs (with the single exception being Ruprecht 106, \citealt{villanova13}). 
This is also found in the Fornax dwarf galaxy \citep{larsen14}, the Sagittarius dwarf galaxy \citep{carretta14}, the WLM dwarf galaxy \citep{larsen14b} and in the Large Magellanic Cloud (LMC, \mbox{\citealt{mucciarelli09}}). Recently, MPs were also detected in the $\sim$ 10 Gyr-old Small Magellanic Cloud (SMC) cluster NGC 121 (\citealt{dalessandro16} and \citealt{paperI}, Paper I hereafter). 

Indeed, until recently, MPs were believed to be found only in massive, old clusters, while none were present in clusters of comparable age and lower masses (e.g., E3, \citealt{salinas15}). Such evidence led many to consider mass as the key cluster property controlling the presence of MPs (e.g., \citealt{gratton12}). The discovery of relatively young ($\sim 1-2$ Gyr old), but still massive ($\lesssim 2 \times 10^5$ \msun) clusters with no abundance spreads within them (\citealt{mucciarelli08}, \citeyear{mucciarelli14}, \citealt{colucci12}) challenged this scenario. In addition to this, \citet{mucciarelli11} analysed the spectra of red giant branch (RGB) stars in NGC 1866, a very young ($\sim 200$ Myr) and massive ($\sim 1 \times 10^5$ \msun) cluster, finding no evidence for MPs. Although these results were based on a relatively small sample of stars with spectroscopically determined abundances, such findings have opened a new question about whether age could be considered as a relevant factor for the existence of MPs, as well as mass. The goal of our ongoing HST survey is to provide new insights into this phenomenon. 

\citet{hollyhead17} spectroscopically detected MPs in the SMC cluster Lindsay 1 which is massive ($\sim 10^5$ \msun) and relatively young ($\sim 8$ Gyr old). \cite{paperII} (Paper II hereafter) photometrically confirmed the result by \citet{hollyhead17} for Lindsay 1 and showed that MPs are present in two other intermediate-age SMC clusters, namely NGC 339 and NGC 416, similar in mass and age to Lindsay 1. Also, Hollyhead et al. (in prep.) have detected MPs in Kron 3, another SMC cluster which is $\sim 6.5$ Gyr old and as massive as $\sim 2 \times 10^5$ \msun.
All these massive clusters fall in a range of ages (i.e., $6-8$ Gyr) which have never been explored. So far, Lindsay 1, NGC 339, NGC 416 and Kron 3 represent the youngest clusters that have been found to host MPs. This surprising result suggests that the MPs phenomenon is not a cosmological effect, as it operated at least down to $z=0.65$, well past the peak
epoch of GC formation ($z \gtrsim 2$, e.g. \citealt{brodie06}).

Here we present the analysis of NGC 419, a young ($\sim 1.5$ Gyr, \citealt{glatt08}), massive ($\gtrsim 2 \times 10^5$ \msun, \citealt{goudfrooij14}) SMC cluster, one of the targets of our photometric survey of Magellanic Clouds (MCs) star clusters (Paper I \& II).
So far in our survey, we detected MPs in the SMC clusters NGC 121 (Paper I), NGC 339, NGC 416 and Lindsay 1 (Paper II). Here we will show that there is no evidence for multiple sequences/broadening in the RGB of NGC 419. 

NGC 419 shows one of the most extended main sequence turnoffs (eMSTOs) in clusters analysed so far, a
common peculiarity of intermediate and young age MC massive clusters, from $1-2$ Gyr (e.g. \citealt{mackey08}, \citealt{milone09}) down to $\sim 100$ Myr (\citealt{bastian16}). Originally, an eMSTO was thought to be caused by age spreads of $\sim 200-700$ Myr (e.g. \citealt{goudfrooij14}) due to multiple star formation (SF) events. According to this hypothesis, NGC 419 should have a very large age spread, $\sim 700$ Myr \citep{rubele10}. However, subsequent works have shown that such features seem more likely due to a single age population with a range of stellar rotation rates (\citealt{bastiandemink09}, \citealt{niederhofer15}, \citealt{brandt15}).

This paper is organised as follows: in \S \ref{subsec:dataphot} and \S \ref{subsec:dr} we describe the observation of NGC 419 and data reduction procedures, while in \S \ref{subsec:analysis} we report the analysis of the cluster. In \S \ref{sec:models} we give an outline about the stellar evolution models we used and in \S \ref{sec:results} we present our results of comparison between data and models. Finally, we discuss our results and conclusions in \S \ref{sec:discussion}.  

\begin{figure}
	\centering
	\includegraphics[scale=0.52]{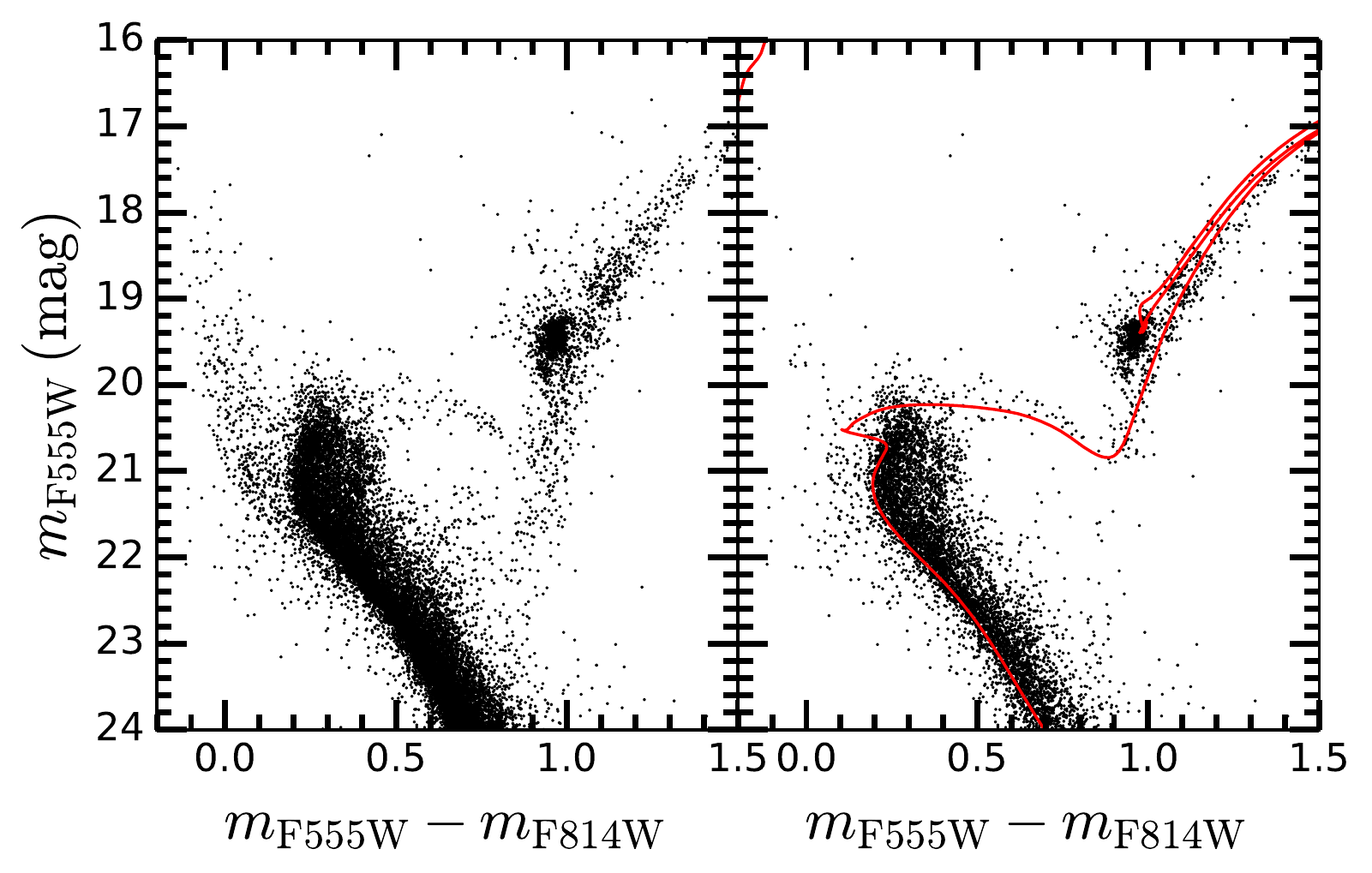}	
	\caption{$m_{F555W}-m_{F814W}$ vs. $m_{F555W}$ CMD of NGC 419 before (left) and after (right) the field star subtraction. The red curve in the right panel indicates the MIST isochrone we adopted for values of age (1.4 Gyr) and metallicity ([Fe/H]${=-0.7}$).}
	\label{fig:contamination}
\end{figure}

\begin{figure*}
	\centering
	\includegraphics[scale=0.46]{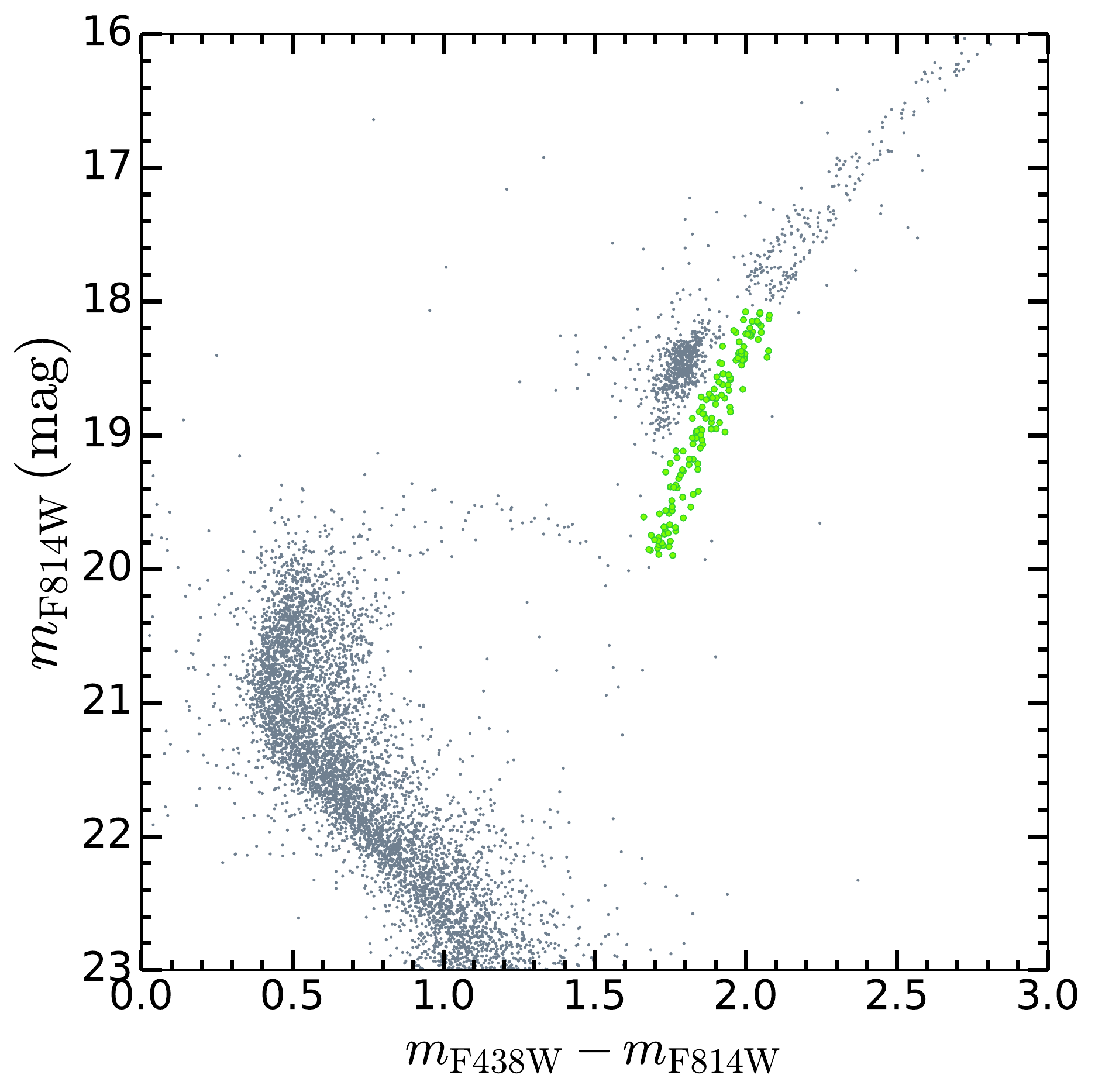}
	\includegraphics[scale=0.46]{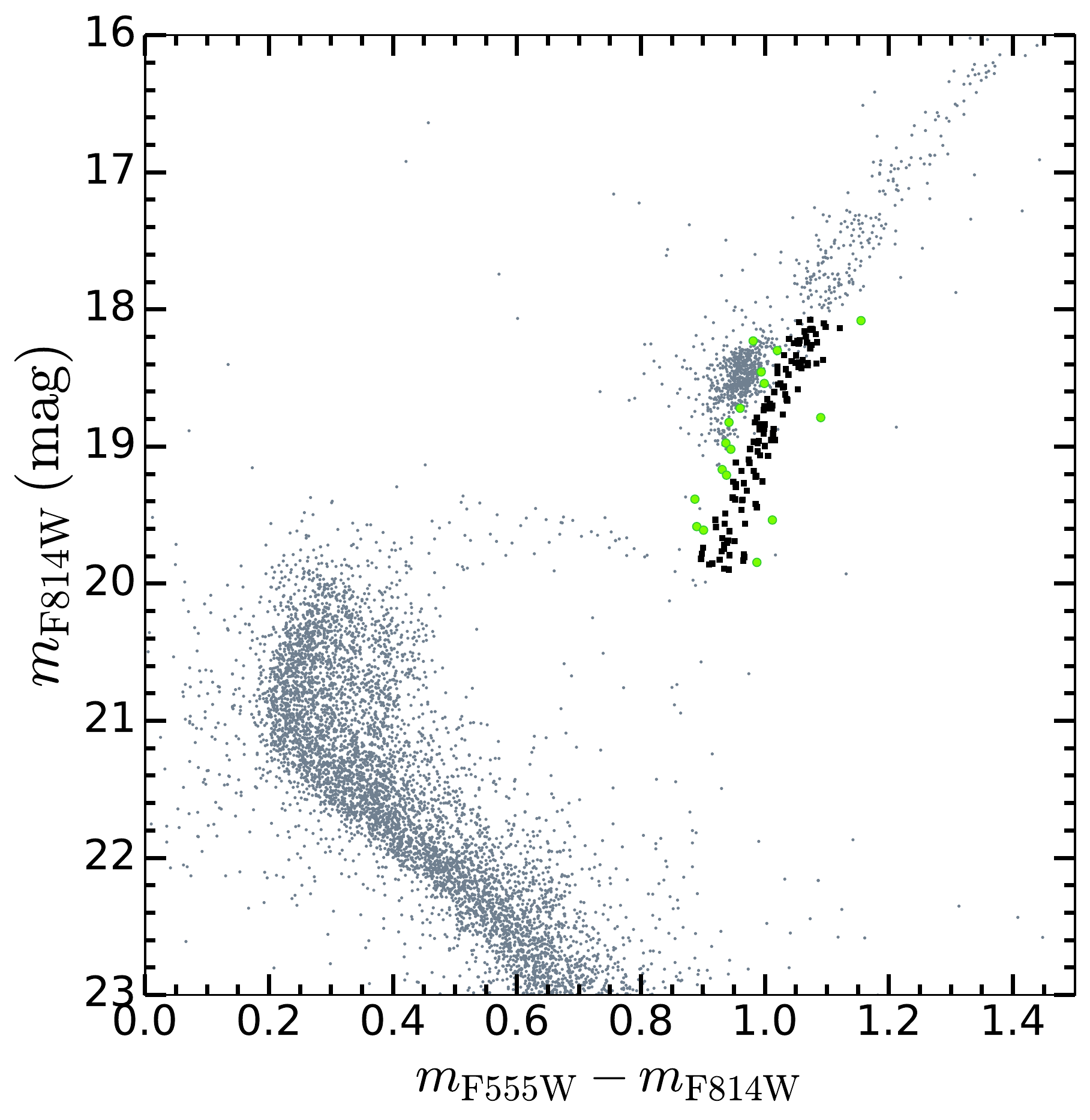}
	\caption{{\it Left panel:} CMD of NGC 419 using $m_{F438W}-m_{F814W}$ vs. $m_{F814W}$. Green filled circles mark the RGB stars selected in this colour and magnitude combination. {\it Right Panel}: $m_{F555W}-m_{F814W}$ vs. $m_{F814W}$ CMD of NGC 419. Black filled squares indicate the second selection of RGB stars in this colour-magnitude space, while green filled circles represent the stars passing the initial selection but not the second.}
	\label{fig:selection}
\end{figure*}

\section{Observations and Data Reduction}
\label{sec:observations}

\subsection{Data and Photometry}
\label{subsec:dataphot}

Photometric data for NGC 419 were obtained from the ongoing \emph{Hubble Space Telescope} survey GO-14069 (P.I. N. Bastian, see Paper I)\footnote{The photometric catalogues are available from the authors upon request.}.
New observations were provided for the optical wide band filter $F438W$ and the UV narrow band filter $F343N$  (WFC3/UVIS instrument). Archival data were used for the UV band filter $F336W$ (WFC3/UVIS, GO-12257, P.I. L.Girardi), and for the optical bands $F555W$ and $F814W$ (ACS/WFC, GO-10396, P.I. J.Gallagher). 

The UV/optical filters are extremely useful when searching for MPs. Strong NH absorption lines are present inside the $F336W$ and $F343N$ filter bands, while the $F438W$ passband includes CH absorption features (\citealt{piotto15} and Fig. 1 in Paper I). 
Both $F555W$ and $F814W$ filters were used to select RGB stars, while the $F814W$ one, along with $F343N$, $F336W$, $F438W$, was also used to compose a suitable colour combination for revealing the presence of MPs (see \S \ref{subsec:analysis}). 

Both ACS/WFC and WFC3/UVIS observations were processed through the standard HST pipeline. The photometry for NGC 419 was determined by 
applying the point spread function (PSF) fitting method, using the spatial variable ``effective PSF'' (ePSF) libraries \citep{anderson06}.
We refer to Paper I and Paper II for more specific details about data processing and photometry.

To select cluster members, we considered stars that are within 900 pixels from the center of NGC 419, which corresponds to 36$''$ (0.04$''$ pixel scale for the WFC3/UVIS instrument). We then defined a background reference region with the same area as the cluster region in order to statistically subtract field stars from the cluster CMD in $m_{F336W}-m_{F438W}$ vs. $m_{F438W}$ space. 
For every star in the background region, the closest star in colour-magnitude space in the cluster region is removed. Fig. \ref{fig:contamination} shows the $m_{F555W}-m_{F814W}$ vs. $m_{F555W}$ CMD of NGC 419, before (left panel) and after (right panel) the field star subtraction.

\subsection{Differential Reddening}
\label{subsec:dr}
We also investigated possible effects of differential reddening on our CMDs, even if the level of extinction towards NGC 419 is quite low ($A_V \simeq 0.15$, \citealt{goudfrooij14}). We corrected the photometry of NGC 419 for differential reddening following the method described in \citet{milone12} and using the extinction coefficients reported in \cite{milone15}. 
The average change in colour due to differential extinction in both $m_{F336W}-m_{F438W}$ and $m_{F343N}-m_{F438W}$ colours results to be $<0.01$, with a maximum $A_{F336W} = 0.02$. We conclude that reddening effects are negligible and do not affect our results. Thus, we did not account for the differential extinction correction in our analysis.

\begin{figure*}
	\centering
	\includegraphics[scale=0.46]{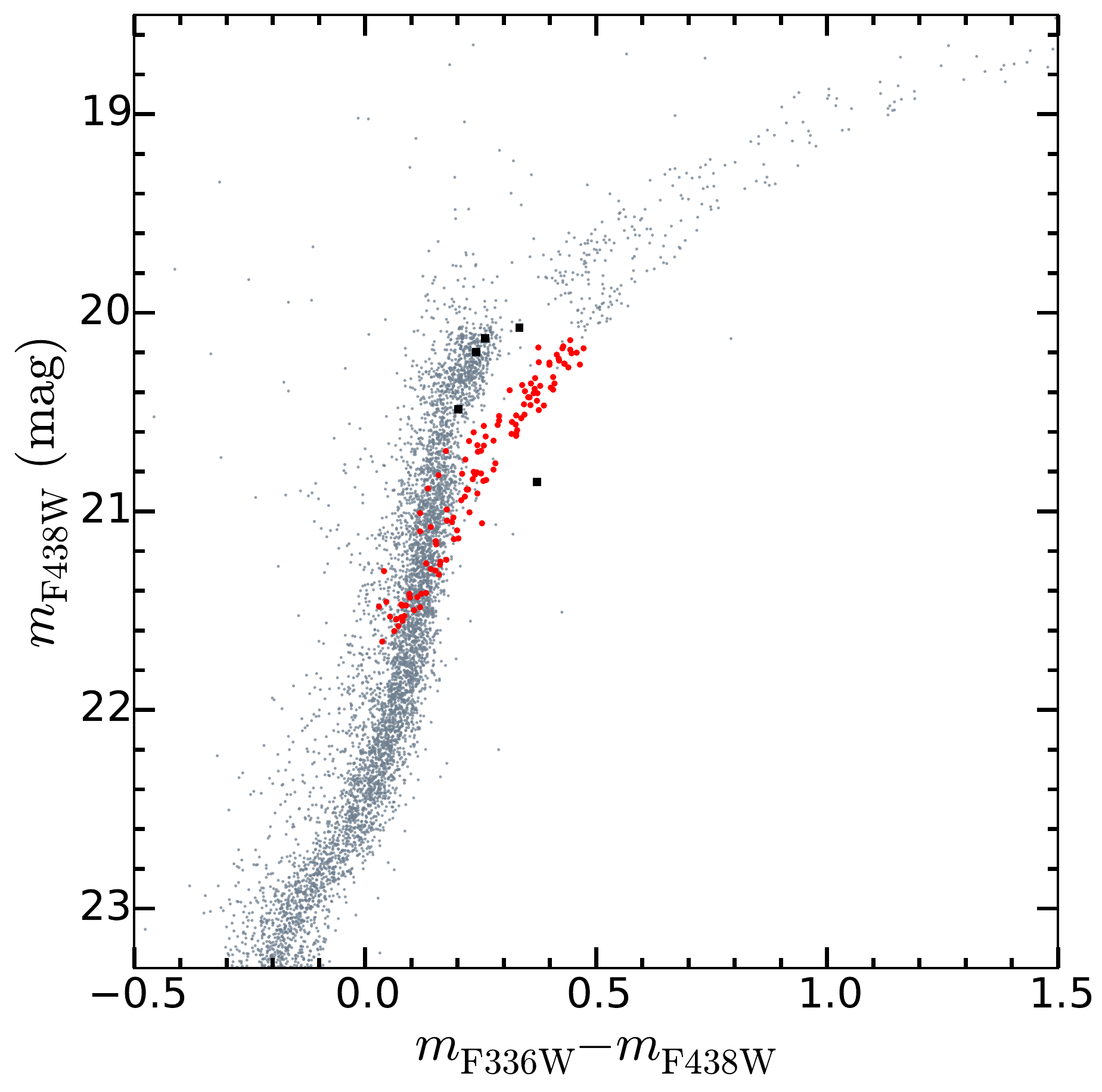}
	\includegraphics[scale=0.46]{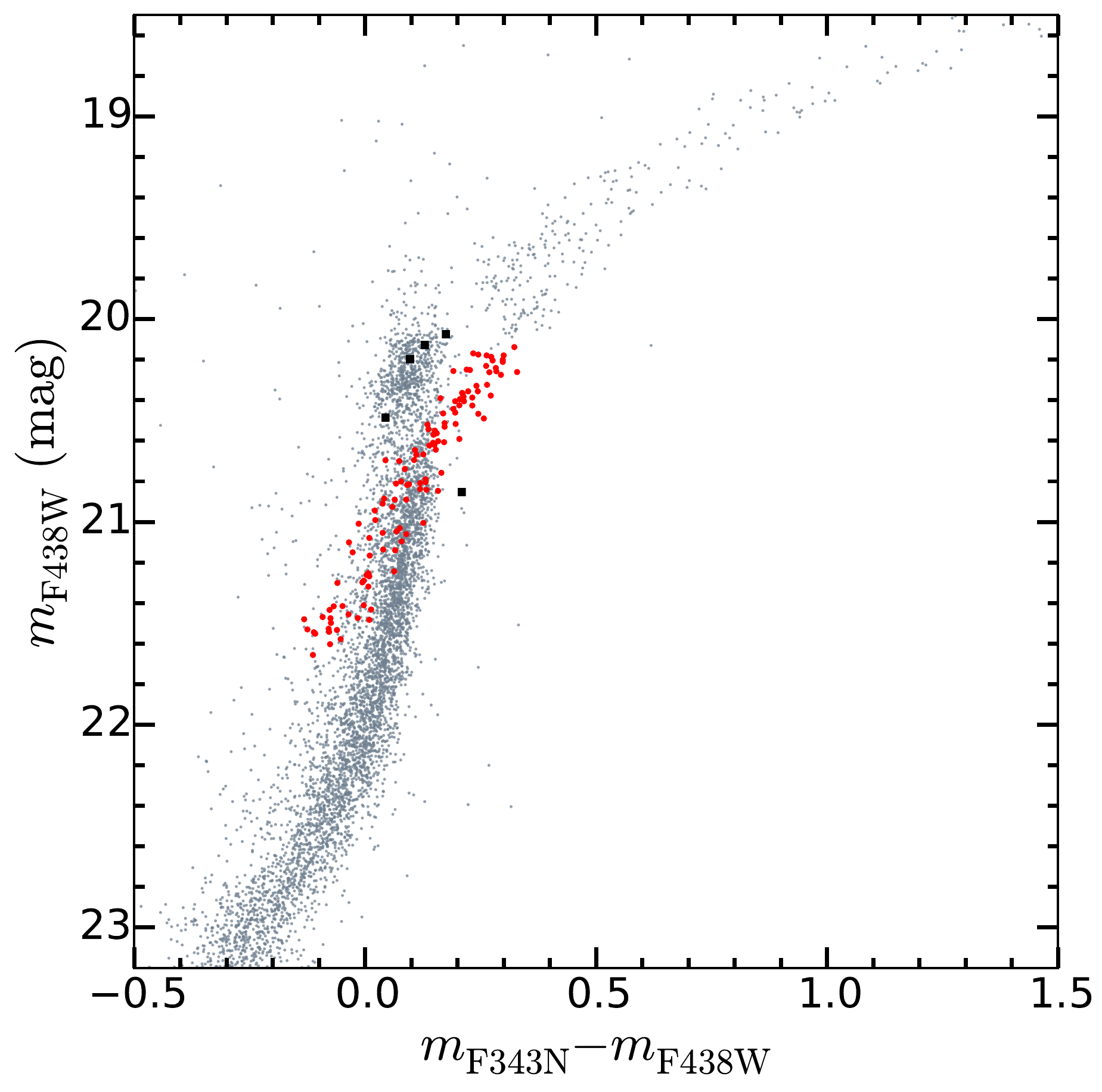}
	\caption{$m_{F336W}-m_{F438W}$ vs. $m_{F438W}$ CMD (left panel) and $m_{F343N}-m_{F438W}$ vs. $m_{F438W}$  CMD (right panel) of NGC 419. Red filled circles indicate the final RGB selected stars, while black filled squares represent the stars which did not pass the final selection.}
	\label{fig:UBselection}
\end{figure*}

\subsection{Analysis}
\label{subsec:analysis}

In order to search for the presence of multiple populations, we need to select a clean sample of RGB cluster members. To be as conservative as possible, we selected RGB stars in three different colour-magnitude spaces. This reduces the contamination by SMC field stars with ages $\sim$ 1 Gyr, comparable to the age of NGC 419. The first selection was made in $F438W-F814W$ colours. Indeed, these colours are the best at separating the RGB from the Asymptotic Giant Branch (AGB) and the Horizontal Branch (HB). 
Optical colours are much less affected by sensitive star-to-star N variations than other colour combinations with a passband encompassing the NH and CN molecular features, i.e. the F336W and F343N filters \citep{sbordone11}. 
The left panel of Figure \ref{fig:selection} shows the $m_{\rm F438W}-m_{\rm F814W}$ vs. $m_{\rm F814W}$ CMD of NGC 419. Green filled circles mark the RGB stars selected in this filter combination. To avoid contamination by AGB stars, a brightness cut  was applied ($m_{\rm F814W} > 18$). We then plotted these RGB stars on the $m_{\rm F555W}-m_{\rm F814W}$ vs. $m_{\rm F814W}$ CMD of NGC 419. We noticed that several objects were scattered off the RGB or found to belong to the red clump, hence we made a second selection in $F555W-F814W$ colours.
The right panel of Fig. \ref{fig:selection} shows the $m_{\rm F555W}-m_{\rm F814W}$ vs. $m_{\rm F814W}$ CMD of NGC 419 with black filled squares indicating the second selection of RGB stars. Green filled circles mark the stars passing the first selection criterion but not the second. Also note that Fig. \ref{fig:selection} shows NGC 419's remarkable eMSTO.

We then plotted the second selection of RGB stars on the $m_{\rm F336W}-m_{\rm F438W}$ vs. $m_{\rm F438W}$ CMD. Again, we found a very few objects which scattered off the RGB or on the tip of the MS (5 out of $>100$ stars). 
We made the final selection in $F336W-F438W$ colours and this is shown in the left panel of Fig. \ref{fig:UBselection}, where the $m_{\rm F336W}-m_{\rm F438W}$ vs. $m_{\rm F438W}$ CMD of NGC 419 is displayed. The right panel of Fig. \ref{fig:UBselection} shows the CMD of NGC 419 using the narrow band filter F343N, in the $m_{\rm F343N}-m_{\rm F438W}$ vs. $m_{\rm F438W}$ space. 
Red filled circles mark the final selected RGB stars in both panels of Fig. \ref{fig:UBselection}. Black filled squares represent the stars that did not pass the third selection. Interestingly, the RGB stars superimpose on the main sequence in these filters, emphasizing the importance of a selection in other colours such as $F438W-F814W$ and $F555W-F814W$.

A first look at the $m_{\rm F336W}-m_{\rm F438W}$ and $m_{\rm F343N}-m_{\rm F438W}$ vs. $m_{\rm F438W}$ diagrams reveals that no splitting is detected in the RGB. The presence of multiple sequences and/or broadening in the RGB is a clear indication of the existence of two or more populations of stars, one with a primordial chemical composition, the others with a certain level of chemical enrichment (depleted in C and O and enhanced in N). 
Accordingly, we performed an analysis in order to quantify the observed spread in the UV/optical CMDs of NGC 419. More specifically, we analysed the differences between the spreads in the two filters, $F336W$ and $F343N$.

\begin{figure}
	\centering
	\includegraphics[scale=0.35]{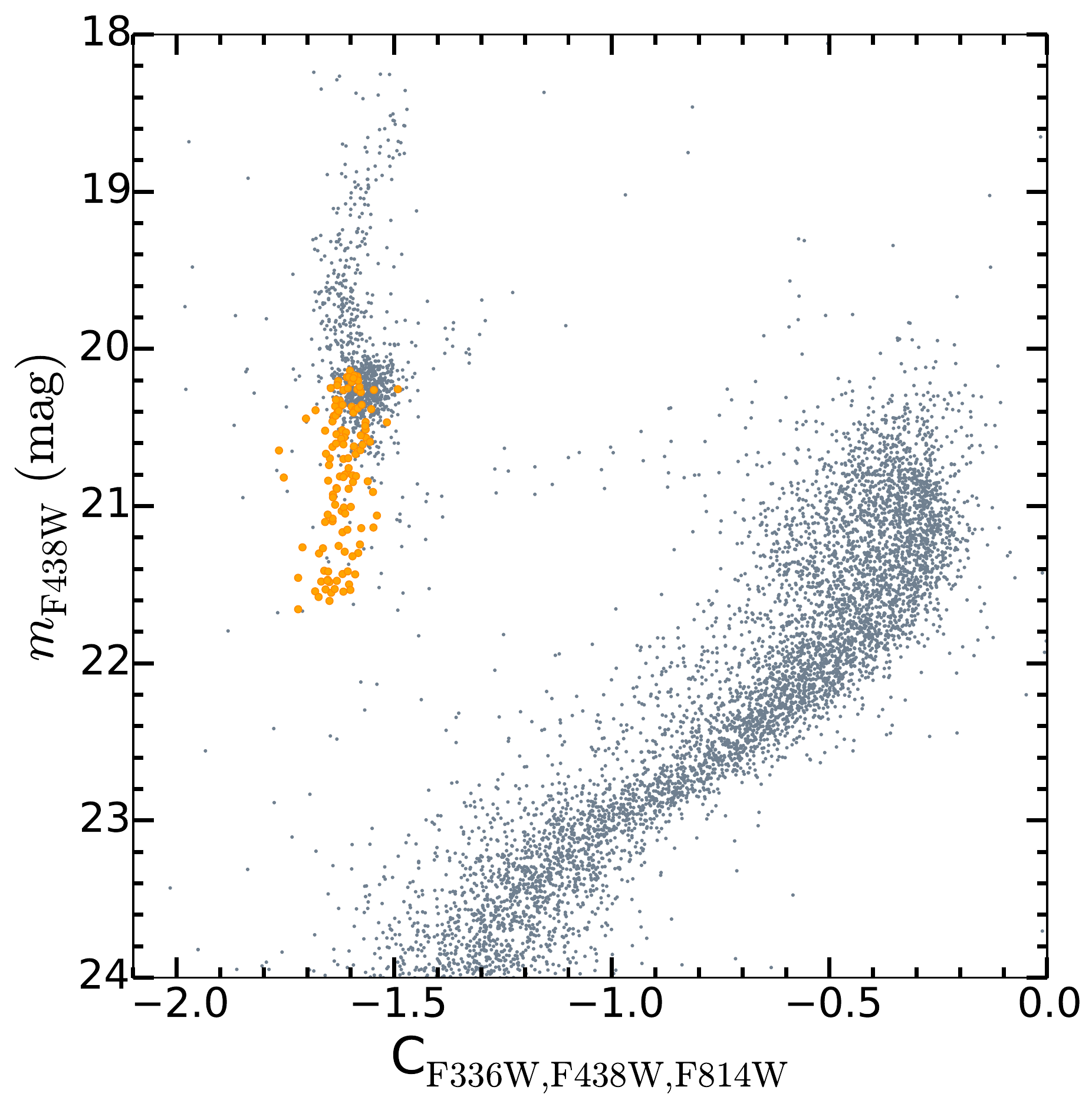}
	\includegraphics[scale=0.35]{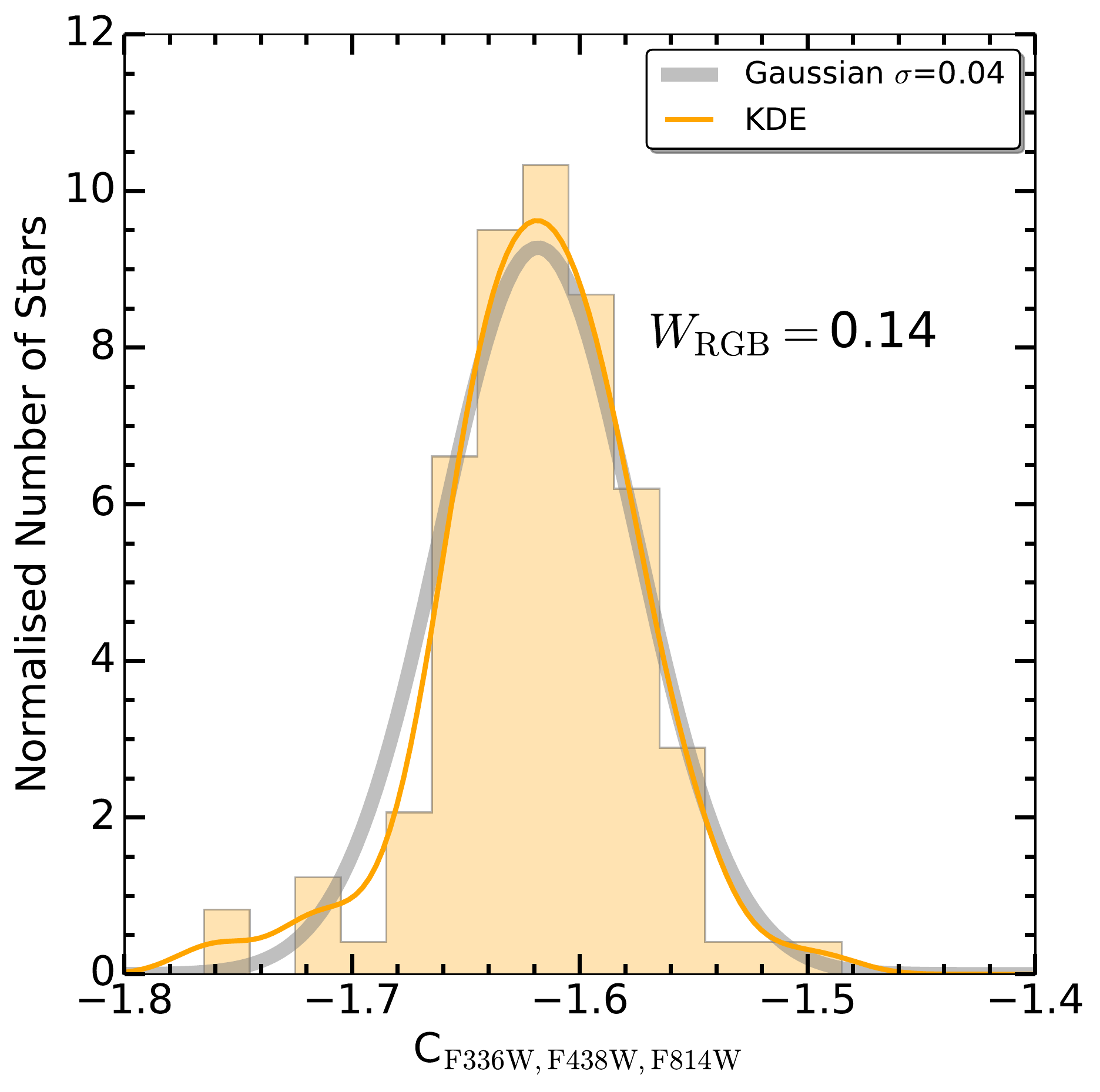}
	\caption{{\it Top Panel}: \cubi  vs. $m_{F438W}$ CMD of NGC 419. Orange filled circles mark the selected RGB stars. {\it Bottom Panel}: histogram of the distribution of RGB stars in NGC 419, in \cubi\ colours. The grey solid curve represents the Gaussian probability density function (PDF) with mean and standard deviation ($\sigma$) calculated on unbinned data, while the orange solid curve indicates the KDE. Superimposed on the plot is the $W_{\rm RGB}$ index, see text for more details.}
	\label{fig:cubi}
\end{figure}

\begin{figure}
	\centering
	\includegraphics[scale=0.35]{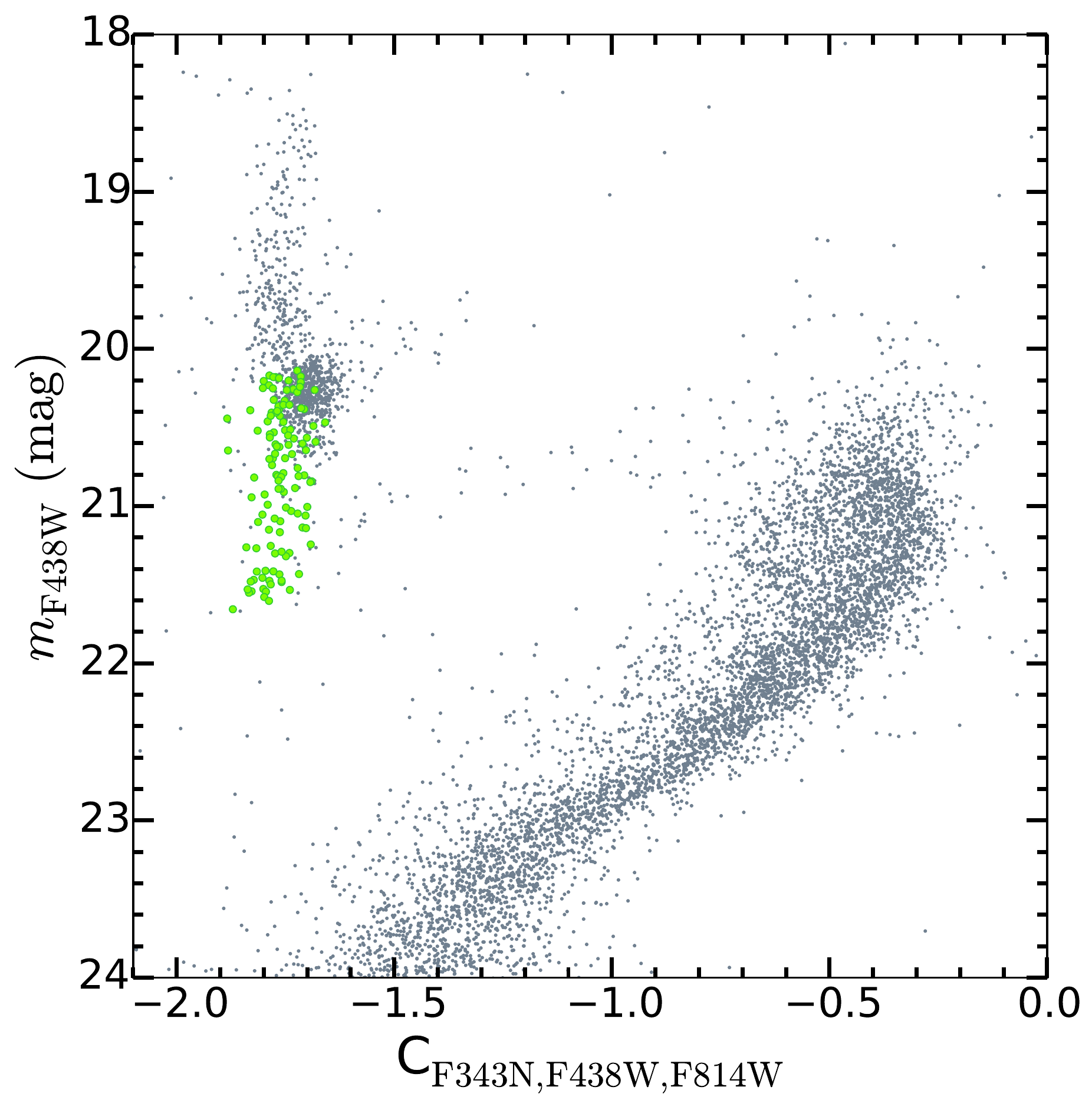}
	\includegraphics[scale=0.35]{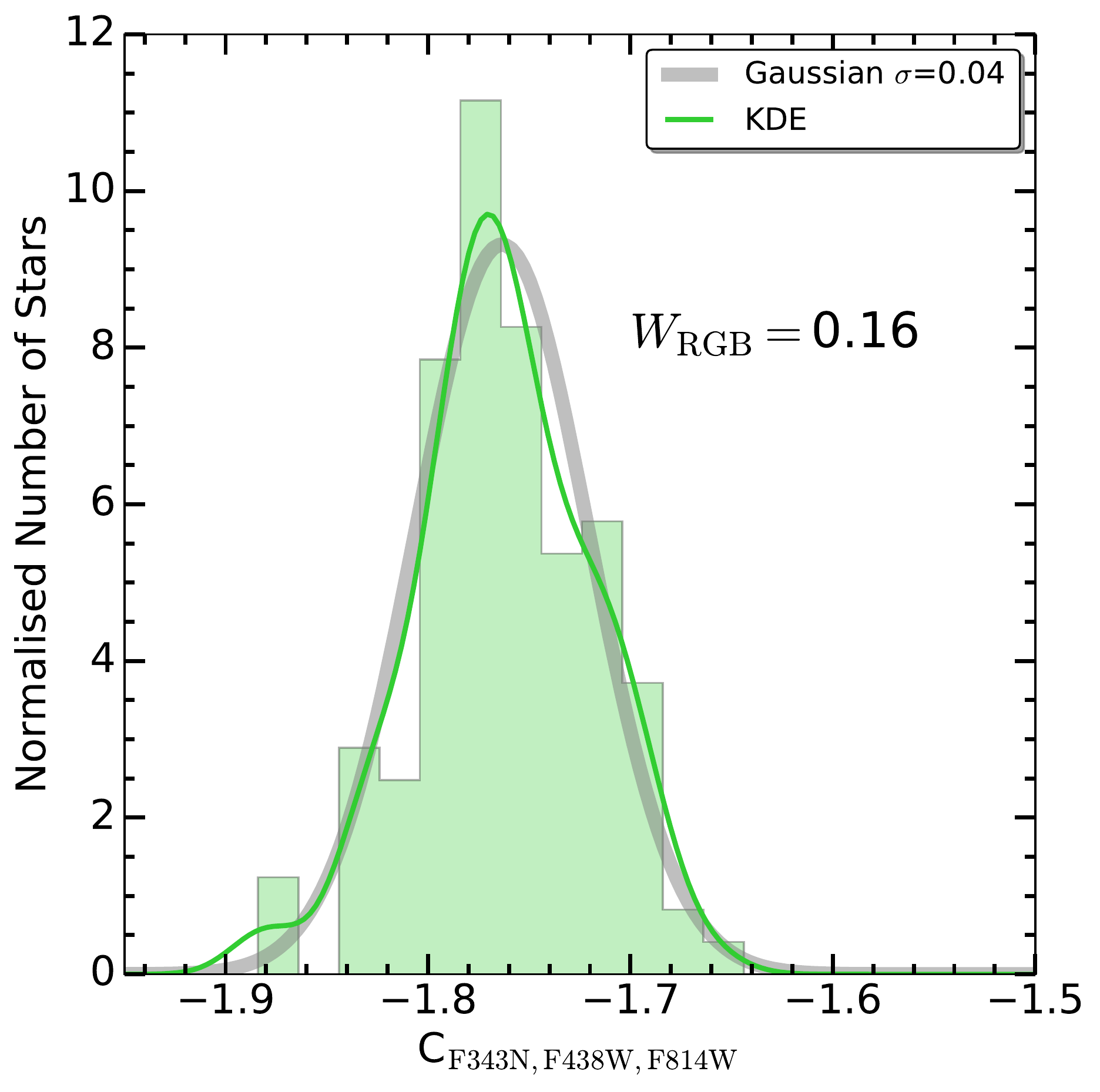}
	\caption{{\it Top Panel}: \cunbi  vs. $m_{F438W}$ CMD of NGC 419. Green filled circles mark the selected RGB stars. {\it Bottom Panel}: histogram of the distribution of RGB stars in NGC 419, in \cunbi\ colours. The grey solid curve represents the Gaussian probability density function (PDF) with mean and standard deviation ($\sigma$) calculated on unbinned data, while the green solid curve indicates the KDE. Superimposed on the plot is the $W_{\rm RGB}$ index, see text for more details.}
	\label{fig:cunbi}
\end{figure}

We used a filter combination of the form $(F336W-F438W)-(F438W-F814W)$ = \cubi\ for the wide band F336W and $(F343N-F438W)-(F438W-F814W)$ = \cunbi\ for the narrow band F343N. \citet{monelli13} used a similar filter combination in order to detect the presence of MPs in a number of Galactic GCs, the $C_{U,B,I} = (U-B)-(B-I)$ combination. They point out that this pseudo-colour is effective at unveiling multiple sequences and spreads in the RGB (see also \S \ref{sec:models} and \S \ref{sec:results} for model predictions). 
Another advantage of this pseudo-colour at the age of NGC 419 is that the RGB is almost vertical. The same colour (\cubi) has also recently been used by \cite{dalessandro16} to efficiently detect MPs in NGC 121. The top panels of Figures \ref{fig:cubi} and \ref{fig:cunbi} show the CMDs of NGC 419 using \cubi\ vs. $m_{F438W}$ and \cunbi\ vs. $m_{F438W}$, respectively.
 Orange and green circles indicate the selected RGB stars in the two different CMDs. No evidence of multiple sequences is seen in such filter combinations either. 
The bottom panels of Figures \ref{fig:cubi} and \ref{fig:cunbi} show the histograms of the distributions in \cubi\ and \cunbi\ colours of the RGB stars in NGC 419, respectively. Hereafter, we will refer to \cubi\ as \cubifancy\ and to \cunbi\ as \cunbifancy\ for more clarity, unless stated otherwise. 

We calculated the mean and standard deviation ($\sigma$) on unbinned colours (i.e., \cubifancy\ and \cunbifancy) and derived a Gaussian PDF, indicated as a grey curve in the bottom panels of Fig. \ref{fig:cubi} and \ref{fig:cunbi}. 
The obtained $\sigma$ values are $\simeq 0.04$ for both filter combinations, with a difference of only $\sim 2 \times 10^{-4}$.
We calculated the statistical error on $\sigma$. Using a bootstrap technique based on 10000 realisations, we found that $\sigma_{\rm CUBI}= 0.043 \pm 0.004$ and $\sigma_{\rm CUnBI} = 0.043 \pm 0.003$. As far as photometric errors are concerned, these are essentially the same in F336W and F343N filters in this bright regime. Therefore, we can say that errors are the same for both \cubifancy\ and \cunbifancy. Hence, the observed RGB widths in \cubifancy\ and \cunbifancy\ colours are directly comparable. In addition to this, the photometric errors in \cubifancy\ and \cunbifancy\ colours are comparable to the observed spreads.

We derived the kernel density distribution (KDE) from a Gaussian kernel for both \cubifancy\ and \cunbifancy\ colours. The results are superimposed on the histograms of data in the bottom panels of Fig. \ref{fig:cubi} and \ref{fig:cunbi}, as orange and green solid curves respectively. By visual inspection, we were unable to detect any significant difference between the gaussian and KDE distributions. The KDE did not reveal any bimodality or peaks that the gaussian could have smoothed out. Indeed, the dip test for unimodality \citep{hartigan85} confirms that there is no statistically significant bimodality in either the \cubifancy\ or \cunbifancy\ distribution.

We also provided a different estimate for the RGB width. We defined the \wrgb\ index as the colour extension of the KDE at 20\% of the distribution maximum. The values of \wrgb\ for \cubifancy\ and \cunbifancy\ are given in the bottom panels of Figs. \ref{fig:cubi} and \ref{fig:cunbi}, respectively.
We estimated the error on the \wrgb\ index using a Monte Carlo simulator technique. We obtained \wrgb$=0.139 \pm 0.013$ for \cubifancy\ and \wrgb$=0.157 \pm 0.013$ for \cunbifancy. This confirms that the two widths are comparable, at $\sim 1\sigma$ confidence level.

Synthetic spectra which take into account the presence of multiple population, i.e. the presence of enriched stars with respect to the primordial ones, predict a significant difference in the observed RGB spread when using the wide-band $F336W$ filter with respect to the narrow-band $F343N$ filter (see \S \ref{sec:models}). More specifically, we observe that $\sigma_{\rm CUnBI}\simeq \sigma_{\rm CUBI} \simeq 0.04$ but we would have expected $\sigma_{\rm CUnBI} \gtrsim 0.1$ if MPs were present (\S \ref{sec:results}). If the spreads are caused only by photometric errors, then we would expect $\sigma_{\rm CUBI} \simeq \sigma_{\rm CUnBI}$, which is what we observe. No clear difference is detected in the observed spreads and \wrgb\ indices between \cubifancy\ and \cunbifancy\ colours, hence we do not detect multiple populations in the RGB of NGC 419. We will discuss in detail these outcomes in Section \ref{sec:results}.

Finally, we also looked at observed spreads in the Red Clump (RC) and upper RGB (URGB). We performed the same analysis as for the RGB. We obtained that the widths are the same for \cubifancy\ and \cunbifancy\ in the RC (\wrc$\simeq 0.12$). We observe a slightly larger spread for \cunbifancy\ with respect to \cubifancy\ in the URGB (\wagb(CUBI)$=0.105 \pm 0.012$, \wagb(CUNBI)$=0.129 \pm 0.016$), although these are comparable at $\sim 1 \sigma$ level when taking the error into account.

\section{Models}
\label{sec:models}

In this Section, we report on how we calculated stellar evolution models with different levels of chemical enrichment, in order to compare them with our observations. We describe the estimation of age and metallicity in \S \ref{subsec:params} and we outline the calculation of models in \S \ref{subsec:anomalies}. In \S \ref{subsec:he}, we consider possible effects of an enhanced He abundance on the RGB.

\subsection{Age, Extinction and Distance Modulus}
\label{subsec:params}

We used MIST isochrones \citep{dotter16, choi16} for several ages ($\log[\rm t/yr] = 9.1, 9.15, 9.2, 9.25$) and metallicities ([Fe/H]$=-0.6, -0.7, -0.75$). We then assumed the distance modulus $M-m = 18.85$ and extinction value $A_V = 0.15$ for NGC 419 from \citet{goudfrooij14}, in order to match data to the isochrones.  For the other filter extinction values, we used $A_{\rm F336W} = 1.64 \, A_V$, $A_{\rm F343N} = 1.64 \, A_V$, $A_{\rm F438W} = 1.35 \, A_V$, $A_{\rm F555W} = 1.055 \, A_V$, $A_{\rm F814W} = 0.586 \, A_V$ \citep{milone15, goudfrooij09} . We selected $\log[\rm t/yr] = 9.15$ ($t \simeq 1.4$ Gyr) and [Fe/H]$=-0.7$ as the parameters which best describe the observed $m_{\rm F555W}-m_{\rm F814W}$ vs. $m_{\rm F814W}$ CMD of NGC 419 by eye. This is in agreement with the results by \citet{goudfrooij14}. In addition to this, we also agree  with \citet{glatt08}, who perform several isochrones fitting, which yield values ranging from 1.2 to 1.6 Gyr. The selected isochrone is shown on the $m_{F555W}-m_{F814W}$ vs. $m_{F555W}$ CMD of NGC 419 in the right panel of Fig. \ref{fig:contamination} as a red curve.

\subsection{Abundance Anomalies}
\label{subsec:anomalies}

We calculated synthetic photometry from model atmospheres with different abundance patterns.
We used version 1.0 of the MIST isochrones \citep{choi16} with an age of 1.41 Gyr and a metallicity of [Fe/H] $= -0.70$ to provide input parameters for our model atmospheres (see \S \ref{subsec:params}).
The 1-D MIST models include a range of physics including diffusion on the MS, rotation in stars more massive than 1.2 M$_{\sun}$, convection including thermohaline and rotational mixing.
We used  \textsc{ATLAS12} \citep{kurucz70, kurucz05} to calculate model atmospheres and \textsc{SYNTHE} \citep{kurucz79, kurucz81} to synthesize spectra.
These models are one-dimensional, static and plane parallel and assume local thermodynamic equilibrium.
We used the same versions of the models used by \citet{sbordone04} and line lists for the atomic data as \citet{larsen12} and \citet{larsen14} who we refer to for further details of our stellar atmosphere calculations.
We also utilise the same \textsc{python} wrappers to \textsc{ATLAS12} and \textsc{SYNTHE} as used by \citet{larsen12}.
For our stellar atmosphere calculations we adopted the \citet{asplund09} solar abundances which are the same as adopted by the MIST models.
For each set of models, we calculated 57 model spectra between 0.7 \msun\ on the MS and the tip of the RGB.
We selected the input masses to calculate model atmospheres by eye in $\log L$-$\log T_{\rm eff}$ space with denser sampling during stellar evolutionary phases such as the MSTO and the base of the RGB where the isochrone displays greater curvature.

We calculated the models using three chemical mixtures.
First, we calculated a set of scaled solar models ([C/Fe] = [N/Fe] = [O/Fe] $= 0$).
Next, we calculated a set of intermediate N-enhancement models with [C/Fe] = [O/Fe] $= -0.1$ and [N/Fe] $= +0.5$.
Lastly, we calculated a set of enriched N-enhancement models with [C/Fe] = [O/Fe] $= -0.6$ and [N/Fe] $= +1.0$.
For the enhanced models, the C and O abundances were chosen to keep the [(C+N+O)/Fe] the same between the models, according to what we observe in standard GCs \citep{brown91, cohen05, yong15, marino16}. 
For each of these chemical mixtures we kept the helium abundance (surface Y $= 0.248$) constant and all other abundances fixed at solar.
We assumed that the model atmospheres had the same chemical abundances at all stellar evolutionary stages.

To produce synthetic magnitudes, we integrated our model spectra over the filter transmission curves for  WFC3\footnote{\url{http://www.stsci.edu/hst/wfc3/ins\_performance/throughputs/Throughput\_Tables}} and ACS/WFC\footnote{\url{http://www.stsci.edu/hst/acs/analysis/throughputs}}.
We then used the zeropoints provided on each instrument's website to calculate Vega magnitudes.
We find excellent agreement ($< 0.01$ mag difference) between our scaled solar models and the photometry calculated by \citet{choi16}.

We then used the synthetic magnitudes to find a suitable combination of colours for revealing the presence of chemical enhancement on the RGB of NGC 419. After trying several filters, we found out that using \cubifancy\ and \cunbifancy\ colours (see \S \ref{subsec:analysis}) appeared to be the most effective way to prove whether NGC 419 showed MPs. Accordingly, when comparing the expected spread in the RGB between the solar and intermediate models, as well as between the solar and enriched models, we obtained that $\Delta($\cunbifancy$) \simeq 2 \times \Delta($\cubifancy$)$. 
This result is directly comparable with the data. We make comparisons with the observed spreads and discuss the outcomes in Section \ref{sec:results}.

\subsection{He Variations}
\label{subsec:he}

We investigated possible effects of an enhanced He abundance on the RGB. Since the MIST isochrones are only available for one He abundance at a given metallicity, we used Padova isochrones \citep{bertelli08} to perform our stellar atmosphere calculations. We assumed the same age and metallicity as in \S \ref{subsec:params} and two different He abundances, one with Y$=0.25$ and one with an enhanced Y$=0.30$. Synthetic photometry was calculated as in Section \ref{subsec:anomalies}, adopting the same solar and enriched chemical mixtures.

We obtained that the difference between the standard and enhanced He models in \cubifancy\ and \cunbifancy\ colours resulted to be $\sim$ 0.01, in the same direction for both solar and enriched mixtures. Hence, these colours are not sensitive to He variations in the RGB and we did not account for enhanced He in our analysis.

\begin{figure}
	\centering
	\includegraphics[scale=0.38]{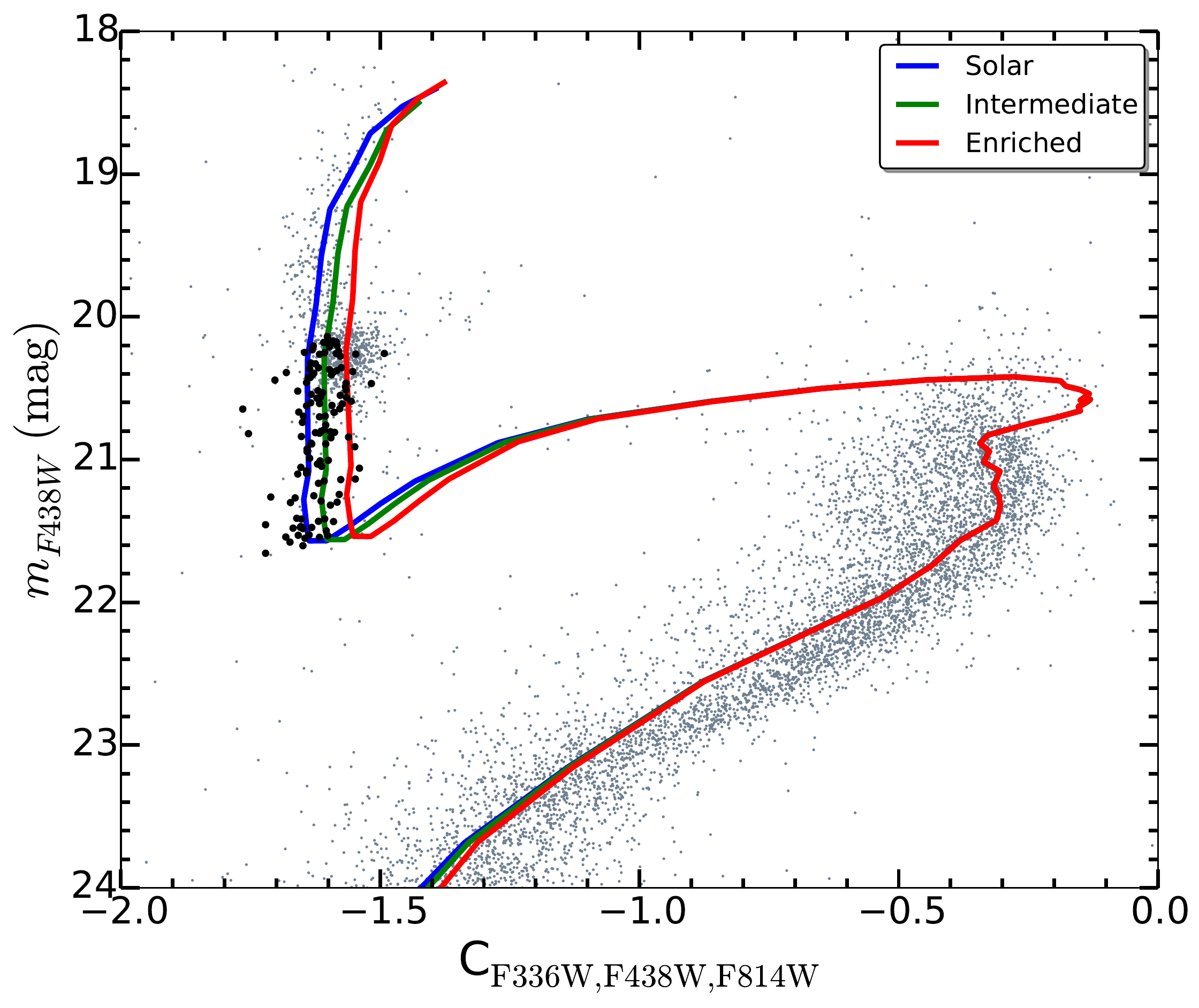}
	\caption{\cubi\  vs. $m_{F438W}$ CMD of NGC 419. Black circles indicate the selected RGB stars. The blue, green and red solid curves represent stellar evolution models ($\log[ {\rm age/Gyr}]= 9.15$, [Fe/H]$=-0.7$) for solar ([C/Fe] = [N/Fe] = [O/Fe] = 0), intermediate ([C/Fe] = [O/Fe] = -0.1, [N/Fe] = +0.5) and enriched ([C/Fe] = [O/Fe] = -0.6, [N/Fe] = +1.0) abundance variations, respectively.}
	\label{fig:cubidatamodels}
\end{figure}

\section{Results}
\label{sec:results}
\begin{figure}
	\centering
	\includegraphics[scale=0.38]{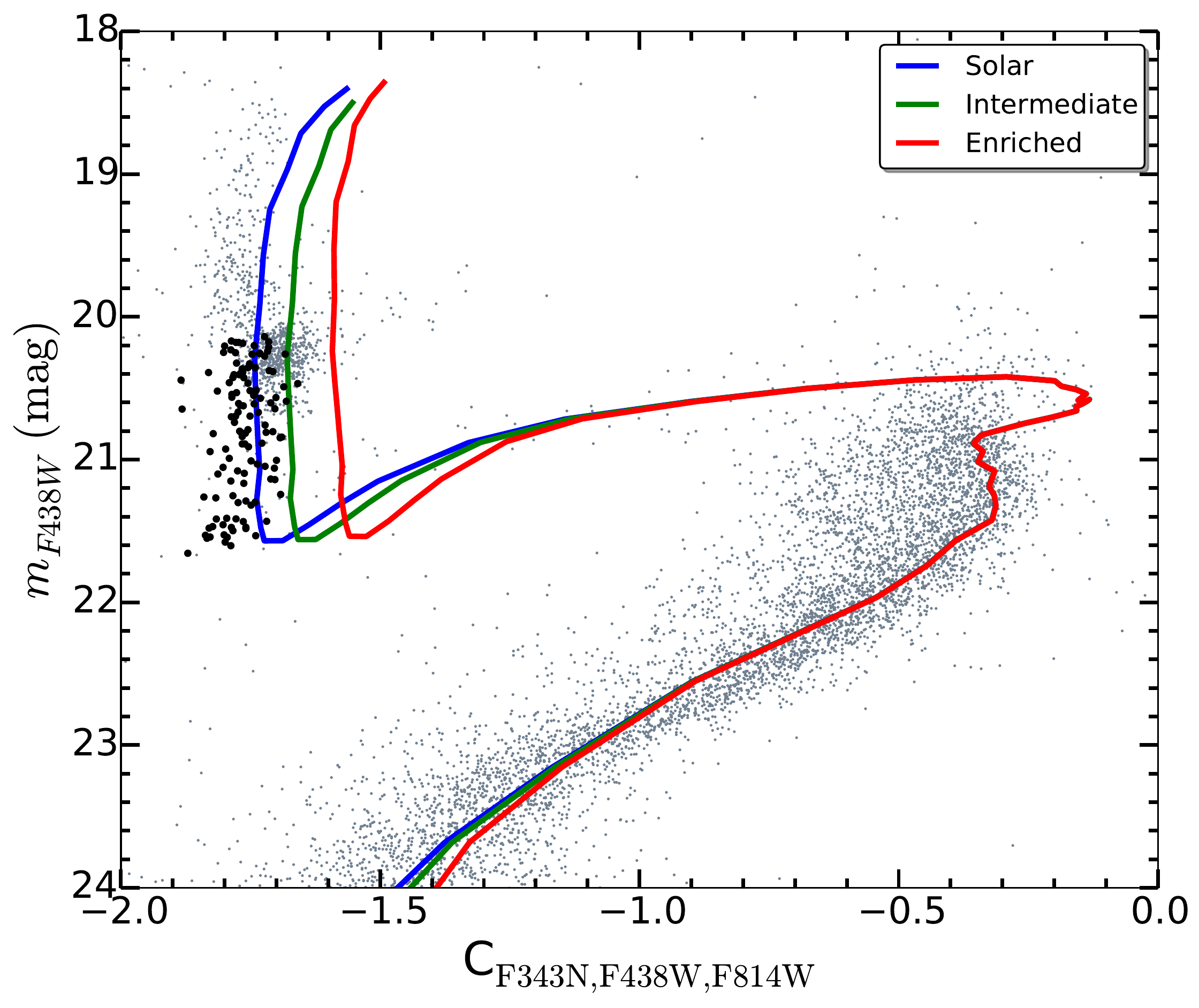}
	\caption{As in Fig. \ref{fig:cubidatamodels} but for \cunbi\ vs. $m_{F438W}$ CMD.}
	\label{fig:cunbidatamodels}
\end{figure}

We compare here our model colours to the data. 
Figures \ref{fig:cubidatamodels} and \ref{fig:cunbidatamodels} show the \cubifancy\  vs. $m_{F438W}$ and \cunbifancy\ vs. $m_{F438W}$ CMDs of NGC 419, respectively, with three different models superimposed. The blue, green and red curves indicate isochrones for solar, intermediate and enriched abundance variations, respectively (see \S \ref{subsec:anomalies}). Black circles indicate the selected RGB stars in both figures. 

According to the models, a first look at Fig. \ref{fig:cubidatamodels} and Fig. \ref{fig:cunbidatamodels} reveals that a difference in the spreads is expected if a chemical variation is present, either intermediate or enriched, between \cubifancy\ and \cunbifancy\ colours. More specifically, we calculated the average spread between the solar and the enriched models in \cubifancy\ colours and this results to be $\simeq 0.079$, while the average spread between these two models in \cunbifancy\ colours is $\simeq 0.156$\footnote{\cubifancy\ and \cunbifancy\ spreads between the solar and enriched model and between solar and intermediate models were calculated in the RGB in a magnitude range $20.2 \leq m_{\rm F438W} \leq 21.2$.}. We then calculated the predicted spread in the RGB from the intermediate enrichment model. While the average spread in the RGB between the solar and the intermediate models in \cubifancy\ colours results to be $\simeq 0.033$, the average one in \cunbifancy\ colours is about $\simeq 0.064$.
Thus, according to the models, the colour spread of stars in the RGB in \cunbifancy\ colours is expected to be twice as broad as the spread in \cubifancy\ colours, if an abundance pattern depleted in C and O and enhanced in N is present (either intermediate or enriched). 

We then compared our predictions to the data. In Section \ref{subsec:analysis}, we have seen that the observed spread of RGB stars in both \cubifancy\ and \cunbifancy\ colours results in $\sigma_{\rm data} \sim 0.04$ ($\sigma_{\rm CUBI} = 0.043 \pm 0.004$ and $\sigma_{\rm CUnBI} = 0.043 \pm 0.003$, see \S \ref{subsec:analysis}). Indeed, these spreads are consistent with that expected from the photometric errors alone. 
We also provided another estimation for the RGB width, \wrgb , and this results in \wrgb$=0.139 \pm 0.013$ for \cubifancy\ and \wrgb$=0.157 \pm 0.013$ for \cunbifancy , Figures \ref{fig:cubi} and \ref{fig:cunbi}.
This proves that we do not observe any significant difference in the RGB spreads between the two colours. Hence, we do not detect the presence of multiple populations either in the form of an enriched ([C/Fe] = [O/Fe] = -0.6, [N/Fe] = +1.0) or intermediate ([C/Fe] = [O/Fe] = -0.1, [N/Fe] = +0.5) chemical anomaly in NGC 419. 
If MPs were present in this cluster, a detection would have been expected by simply comparing the observed width of the RGB in these two colours and by finding a significantly broader $\sigma$ and \wrgb\ (about twice as much) in the narrow-band F343N filter colour combination. We can then conclude that no MPs are detected on the RGB of NGC 419. 
If MPs were present in the way they have been detected in GCs or intermediate-age clusters, we would have been able to observe them in NGC 419 as well. Our analysis can set a limit on [N/Fe] enhancement for NGC 419 to be $<$+0.5 dex, according to what we derive from the intermediate models. However, it is crucial here to state that N enhancements previously observed in intermediate age clusters showing MPs are far higher, e.g. [N/Fe]$>+1.0$ dex for Lindsay 1\citep{hollyhead17}.

Lastly, we compared the expected spreads from the models with the observed spreads in the RC and URGB. As described in Sec. \ref{subsec:analysis}, the observed RC width in \cubifancy\ is as large as the one in \cunbifancy.
From the models we would expect the \cubifancy\ spread to be as twice as much as the \cunbifancy\ one, if a intermediate or enriched chemical variation was present. 
Concerning the URGB, despite the fact that the predicted spreads in the URGB are slightly smaller than in the RGB, we still obtain a ratio between expected \cubifancy\ and \cunbifancy\ colour spreads of $\sim 2$.  We do observe a slightly larger \wagb\ in \cunbifancy\ than in \cubifancy , 
however they are comparable within the errors, at $\sim 1 \sigma$ confidence level (\wagb(CUBI)$=0.105 \pm 0.012$, \wagb(CUNBI)$=0.129 \pm 0.016$).

\section{Discussion and Conclusions}
\label{sec:discussion}

In this paper, we analysed new and archival HST images of the SMC cluster NGC 419. We selected RGB stars by using three different CMDs and colour combinations ($F438W-F814W$, $F555W-F814W$, $F336W-F438W$). We used the pseudo-colour indices \cubi(\cubifancy) and \cunbi(\cunbifancy) in order to maximise the effects of multiple populations on the CMDs along the RGB. No splittings were detected, specifically in the lower RGB. Hence, we quantified the spreads in \cubifancy\ and \cunbifancy\ colours of RGB stars and compared them: these have resulted to be almost equal for both filters ($\sigma_{\rm data}=0.043$, \wrgb (\cubifancy)$=0.139 \pm 0.013$, \wrgb(\cunbifancy)$=0.157 \pm 0.013$).

We generated MIST isochrones to have an estimate of the age and metallicity of NGC 419 and used these to develop models with different chemical anomalies. A solar, intermediate and enriched levels of enhancement have been adopted for a comparison with data. 
The predicted spread in \cubifancy\ between solar and enriched isochrones was found to be half as broad as the \cunbifancy\ spread. The same outcome is seen in the spreads between solar and intermediate isochrones.

We would have expected a significant variation in the observed spreads between \cubifancy\ and \cunbifancy\ colours if MPs were present in NGC 419. We can firmly conclude that no MPs are detected in the RGB of this young ($\sim 1.4$ Gyr) and massive ($ \gtrsim 2 \times 10^5$ \msun) star cluster. 
According to our analysis, we can put a limit on [N/Fe] enhancement for NGC 419 to be $<$+0.5 dex, which is much lower compared to spreads observed in intermediate age clusters showing the presence of MPs (e.g., [N/Fe]$>$+1.0 dex, Lindsay 1, \citealt{hollyhead17}).

This is not the first work to find a lack of evidence for MPs in clusters younger than $\sim 6$ Gyr. \citet{mucciarelli08}, (\citeyear{mucciarelli11}), (\citeyear{mucciarelli14}) studied six intermediate/young age clusters (namely: NGC 1651, 1783, 1978 and 2173; NGC 1866; NGC 1806) in the LMC and found no significant abundance spreads within them, although for each cluster this result is affected by the low number statistics of stars with spectroscopically determined abundances.

However, it is worth stressing that NGC 419 is the first cluster in our HST photometric survey which does not show evidence for MPs. In Paper I we detected MPs in the SMC cluster NGC 121, while in Paper II we detected MPs in 3 additional SMC clusters, namely Lindsay 1, NGC 339, NGC 416. This has been spectroscopically corroborated by the work by \citet{hollyhead17}, which found abundance variations in Lindsay 1, as well. Moreover, Hollyhead et al. (in prep.) has also found spectroscopic evidence for MPs in one more SMC cluster, Kron 3. All the GCs studied so far in our survey reside in the SMC and they are massive, ranging between $\sim 1-2 \times 10^5$ \msun. Nonetheless, they span a wide range in ages from 1.5 to 10 Gyr. NGC 419 is the youngest, while NGC 121 is the oldest one ($\sim 10 $ Gyr). Lindsay 1, NGC 339, NGC 416 and Kron 3 have intermediate ages (from $\sim 6$ up to $\sim 8$ Gyr). 

Our results show that GC mass can no longer be considered as the only key physical property in order to regulate the presence of MPs (see also \citealt{cabrera16a}). Other factors might contribute, such as age, which could play a major role in the development of MPs. Indeed, no massive GCs aged less than $\sim 6$ Gyr have been found with chemical spreads so far. However, this would not be universal, since many less massive, Galactic open clusters older than 6 Gyr also do not host MPs (see the recent compilation by \citealt{krause16}). 

We can estimate the amount of mass loss that NGC 419 will undergo over the next 4.5 Gyr (i.e. from its current age of ~1.5 Gyr to an age of 6 Gyr, where clusters are observed to host MPs - \citealt{hollyhead17}; \citealt{paperII}).  
In order to estimate this we use the rotation curve of the SMC measured by \citet{stanimirovic04} and extrapolate the observations to the galactocentric distance of NGC 419, namely $\sim10$ kpc \citep{glatt08}, obtaining an estimate of $60-70$ km/s.  We also assume that the cluster is tidally filling (in order to maximise the stellar mass loss).  We apply equation 7 of \citet{kruijssen09} (see also \citealt{BM03}, equation 10 and \citealt{lamers05}) to find the dissolution timescale normalisation, $t_{\rm 0}$.  Applying this normalisation to the mass of NGC 419 ($\sim2\times10^5$ \msun) to find the dissolution timescale, $t_{\rm dis} = t_{\rm 0}*M^{\gamma}$ (adopting $\gamma=0.62$ - e.g., \citealt{kruijssen09}), we find a $t_{\rm dis, NGC419} = 152$ Gyr.  If we assume that the cluster loses mass linearly (see the discussion in \citealt{lamers05}) we find that over the next 4.5 Gyr the cluster will lose $\sim3$\% of its mass\footnote{For this, we assumed a circular orbit around the SMC for NGC 419. If we assume a high eccentricity orbit, the mass loss rate due to dissolution could be up to a factor of $\sim$2 higher.}.  Additionally, the cluster is expected to lose of order $9$\% of its mass due to stellar evolution.  NGC 419 is expected to lose, in total,  roughly $\sim12$\% of its current mass over the next 4.5 Gyr (this is an upper limit as we assumed it was tidally limited).  Hence, we conclude that NGC 419 (similarly to the other clusters in our sample in the SMC/LMC, with $M\gtrsim10^5$ \msun) is not expected to undergo significant mass loss over the next few Gyr (see also \citealt{cabrera16}).

In addition to this, in the same galaxy (i.e., the SMC), we have found both the presence and absence of the MPs phenomenon. \citet{glatt08} reported the distribution of star clusters in the SMC, by using distances derived from isochrones fitting. We noticed that our sample of clusters appears to be distributed over a large range of distances with respect to the galaxy centroid. NGC 339 results to be the closest to the SMC centre ($\sim 0.7 \pm 2.0$ kpc), although with a relatively large uncertainty. This is followed by NGC 416 with a distance of $\sim 4$ kpc, Kron 3 ($\sim 7$ kpc) and NGC 121 ($\sim 9$ kpc). The furthest cluster is Lindsay 1 (more than 13 kpc away), while our cluster, NGC 419, is $\sim 10$ kpc away from the SMC centre. 
Accordingly, it appears that the mechanism responsible for enrichment does not depend strongly on the current environment which surrounds the cluster.

To date, the youngest GCs which show MPs are NGC 416 \citep{paperII} and Kron 3 (Hollyhead et al. in prep). At this age ($\sim$ 6 Gyr) and metallicity ([Fe/H] $\simeq -1$) we are sampling RGB stars with masses of $\sim 1$ \msun, while at the age of NGC 419 ($\sim$ 1.5  Gyr) we search for MPs at stellar masses of 1.6 \msun, in the RGB. Hence, our results might also imply that the MPs phenomenon could manifest only below $\sim$ 1 \msun\ and be also due to a stellar evolutionary effect. However, further investigations are needed to confirm whether this is the case.

As stated in \S \ref{sec:intro}, NGC 419 shows one of the largest eMSTO, well noticeable in the $m_{\rm F438W}-m_{\rm 814W}$ and $m_{\rm F555W}-m_{\rm 814W}$ vs. $m_{\rm 814W}$ CMDs in Fig. \ref{fig:selection} and also in \cubifancy\ and \cunbifancy\ colours (Fig. \ref{fig:cubi} and Fig. \ref{fig:cunbi}). Although this is beyond the scope of this paper, our results place limits on the explanation of the eMSTO feature as an age spread \citep{goudfrooij14}. Indeed, if multiple SF episodes due to gas processed by a first generation of stars occurred within the cluster, this would lead to self-enrichment. We did not observe any chemical spread in NGC 419, hence our data might lend support to alternative interpretations, e.g. that the MS spread is caused by a distribution of rotational velocities (e.g., \citealt{bastiandemink09}; \citealt{brandt15}; \citealt{niederhofer15}; \citealt{wu16}).

The results presented in this paper, along with Paper I \& Paper II, highlight that age could play a decisive role in determining the presence of MPs. On the contrary, mass or environment can be excluded as the only key factors in this scenario. However, a larger sample is needed in order to confirm such trends. We will continue our analysis of Magellanic Cloud clusters aged $\lesssim 1-2$ Gyr whose data are already in hand and present the results in forthcoming papers.

\section*{Acknowledgments}
We thank Diederik Kruijssen and Henny Lamers for constructive and helpful discussions on cluster mass loss.
We, in particular F.N., N.B., I.P. and V. K.-P., gratefully acknowledge financial support for this project provided
by NASA through grant HST-GO-14069 for the Space Telescope Science Institute, which is operated by the Association
of Universities for Research in Astronomy, Inc., under NASA contract NAS526555. N.B. gratefully acknowledges 
financial support from the Royal Society (University Research Fellowship) and the European Research Council
(ERC-CoG-646928-Multi-Pop). We are grateful to Jay Anderson for sharing with us his ePSF software.
D.G. gratefully acknowledges support from the Chilean BASAL Centro de Excelencia en Astrof\'isica
y Tecnolog\'ias Afines (CATA) grant PFB-06/2007.

\bibliographystyle{mn2e}
\bibliography{ngc419.bib}

\begin{thebibliography}{55}
\expandafter\ifx\csname natexlab\endcsname\relax\def\natexlab#1{#1}\fi

\bibitem[{{Anderson} \& {King}(2006)}]{anderson06}
{Anderson} J., {King} I.~R., 2006, {PSFs, Photometry, and Astronomy for the
  ACS/WFC}. Tech. rep.

\bibitem[{{Asplund} {et~al}\mbox{.}(2009){Asplund}, {Grevesse}, {Sauval}, \&
  {Scott}}]{asplund09}
{Asplund} M., {Grevesse} N., {Sauval} A.~J., {Scott} P., 2009, ARA\&A, 47, 481

\bibitem[{{Bastian} \& {de Mink}(2009)}]{bastiandemink09}
{Bastian} N., {de Mink} S.~E., 2009, MNRAS, 398, L11

\bibitem[{{Bastian} {et~al}\mbox{.}(2016){Bastian}, {Niederhofer},
  {Kozhurina-Platais}, {Salaris}, {Larsen}, {Cabrera-Ziri}, {Cordero},
  {Ekstr{\"o}m}, {Geisler}, {Georgy}, {Hilker}, {Kacharov}, {Li}, {Mackey},
  {Mucciarelli}, \& {Platais}}]{bastian16}
{Bastian} N. {et~al.}, 2016, MNRAS, 460, L20

\bibitem[{{Baumgardt} \& {Makino}(2003)}]{BM03}
{Baumgardt} H., {Makino} J., 2003, MNRAS, 340, 227

\bibitem[{{Bertelli} {et~al}\mbox{.}(2008){Bertelli}, {Girardi}, {Marigo}, \&
  {Nasi}}]{bertelli08}
{Bertelli} G., {Girardi} L., {Marigo} P., {Nasi} E., 2008, A\&A, 484, 815

\bibitem[{{Brandt} \& {Huang}(2015)}]{brandt15}
{Brandt} T.~D., {Huang} C.~X., 2015, ApJ, 807, 25

\bibitem[{{Brodie} \& {Strader}(2006)}]{brodie06}
{Brodie} J.~P., {Strader} J., 2006, ARA\&A, 44, 193

\bibitem[{{Brown}, {Wallerstein} \& {Oke}(1991){Brown}, {Wallerstein}, \&
  {Oke}}]{brown91}
{Brown} J.~A., {Wallerstein} G., {Oke} J.~B., 1991, AJ, 101, 1693

\bibitem[{{Cabrera-Ziri} {et~al}\mbox{.}(2016{\natexlab{a}}){Cabrera-Ziri},
  {Bastian}, {Hilker}, {Davies}, {Schweizer}, {Kruijssen},
  {Mej{\'{\i}}a-Narv{\'a}ez}, {Niederhofer}, {Brandt}, {Rejkuba}, {Bruzual}, \&
  {Magris}}]{cabrera16}
{Cabrera-Ziri} I. {et~al.}, 2016{\natexlab{a}}, MNRAS, 457, 809

\bibitem[{{Cabrera-Ziri} {et~al}\mbox{.}(2016{\natexlab{b}}){Cabrera-Ziri},
  {Lardo}, {Davies}, {Bastian}, {Beccari}, {Larsen}, \&
  {Hernandez}}]{cabrera16a}
{Cabrera-Ziri} I., {Lardo} C., {Davies} B., {Bastian} N., {Beccari} G.,
  {Larsen} S.~S., {Hernandez} S., 2016{\natexlab{b}}, MNRAS, 460, 1869

\bibitem[{{Carretta} {et~al}\mbox{.}(2014){Carretta}, {Bragaglia}, {Gratton},
  {D'Orazi}, {Lucatello}, \& {Sollima}}]{carretta14}
{Carretta} E., {Bragaglia} A., {Gratton} R.~G., {D'Orazi} V., {Lucatello} S.,
  {Sollima} A., 2014, A\&A, 561, A87

\bibitem[{{Choi} {et~al}\mbox{.}(2016){Choi}, {Dotter}, {Conroy}, {Cantiello},
  {Paxton}, \& {Johnson}}]{choi16}
{Choi} J., {Dotter} A., {Conroy} C., {Cantiello} M., {Paxton} B., {Johnson}
  B.~D., 2016, ApJ, 823, 102

\bibitem[{{Cohen} \& {Mel{\'e}ndez}(2005)}]{cohen05}
{Cohen} J.~G., {Mel{\'e}ndez} J., 2005, AJ, 129, 303

\bibitem[{{Colucci} {et~al}\mbox{.}(2012){Colucci}, {Bernstein}, {Cameron}, \&
  {McWilliam}}]{colucci12}
{Colucci} J.~E., {Bernstein} R.~A., {Cameron} S.~A., {McWilliam} A., 2012, ApJ,
  746, 29

\bibitem[{{Dalessandro} {et~al}\mbox{.}(2016){Dalessandro}, {Lapenna},
  {Mucciarelli}, {Origlia}, {Ferraro}, \& {Lanzoni}}]{dalessandro16}
{Dalessandro} E., {Lapenna} E., {Mucciarelli} A., {Origlia} L., {Ferraro}
  F.~R., {Lanzoni} B., 2016, ApJ, 829, 77

\bibitem[{{Dotter}(2016)}]{dotter16}
{Dotter} A., 2016, ApJS, 222, 8

\bibitem[{{Glatt} {et~al}\mbox{.}(2008){Glatt}, {Grebel}, {Sabbi}, {Gallagher},
  {Nota}, {Sirianni}, {Clementini}, {Tosi}, {Harbeck}, {Koch}, {Kayser}, \& {Da
  Costa}}]{glatt08}
{Glatt} K. {et~al.}, 2008, AJ, 136, 1703

\bibitem[{{Goudfrooij} {et~al}\mbox{.}(2014){Goudfrooij}, {Girardi},
  {Kozhurina-Platais}, {Kalirai}, {Platais}, {Puzia}, {Correnti}, {Bressan},
  {Chandar}, {Kerber}, {Marigo}, \& {Rubele}}]{goudfrooij14}
{Goudfrooij} P. {et~al.}, 2014, ApJ, 797, 35

\bibitem[{{Goudfrooij} {et~al}\mbox{.}(2009){Goudfrooij}, {Puzia},
  {Kozhurina-Platais}, \& {Chandar}}]{goudfrooij09}
{Goudfrooij} P., {Puzia} T.~H., {Kozhurina-Platais} V., {Chandar} R., 2009, AJ,
  137, 4988

\bibitem[{{Gratton}, {Carretta} \& {Bragaglia}(2012){Gratton}, {Carretta}, \&
  {Bragaglia}}]{gratton12}
{Gratton} R., {Carretta} E., {Bragaglia} A., 2012, A\&ARv, 20, 50

\bibitem[{Hartigan \& Hartigan(1985)}]{hartigan85}
Hartigan J.~A., Hartigan P.~M., 1985, The Annals of Statistics, 13, 70

\bibitem[{{Hollyhead} {et~al}\mbox{.}(2017){Hollyhead}, {Kacharov}, {Lardo},
  {Bastian}, {Hilker}, {Rejkuba}, {Koch}, {Grebel}, \&
  {Georgiev}}]{hollyhead17}
{Hollyhead} K. {et~al.}, 2017, MNRAS, 465, L39

\bibitem[{{Krause} {et~al}\mbox{.}(2016){Krause}, {Charbonnel}, {Bastian}, \&
  {Diehl}}]{krause16}
{Krause} M.~G.~H., {Charbonnel} C., {Bastian} N., {Diehl} R., 2016, A\&A, 587,
  A53

\bibitem[{{Kruijssen} \& {Mieske}(2009)}]{kruijssen09}
{Kruijssen} J.~M.~D., {Mieske} S., 2009, A\&A, 500, 785

\bibitem[{{Kurucz}(1970)}]{kurucz70}
{Kurucz} R.~L., 1970, SAO Special Report, 309

\bibitem[{{Kurucz}(2005)}]{kurucz05}
{Kurucz} R.~L., 2005, Memorie della Societa Astronomica Italiana Supplementi,
  8, 14

\bibitem[{{Kurucz} \& {Avrett}(1981)}]{kurucz81}
{Kurucz} R.~L., {Avrett} E.~H., 1981, SAO Special Report, 391

\bibitem[{{Kurucz} \& {Furenlid}(1979)}]{kurucz79}
{Kurucz} R.~L., {Furenlid} I., 1979, SAO Special Report, 387

\bibitem[{{Lamers} {et~al}\mbox{.}(2005){Lamers}, {Gieles}, {Bastian},
  {Baumgardt}, {Kharchenko}, \& {Portegies Zwart}}]{lamers05}
{Lamers} H.~J.~G.~L.~M., {Gieles} M., {Bastian} N., {Baumgardt} H.,
  {Kharchenko} N.~V., {Portegies Zwart} S., 2005, A\&A, 441, 117

\bibitem[{{Larsen} {et~al}\mbox{.}(2014{\natexlab{a}}){Larsen}, {Brodie},
  {Forbes}, \& {Strader}}]{larsen14b}
{Larsen} S.~S., {Brodie} J.~P., {Forbes} D.~A., {Strader} J.,
  2014{\natexlab{a}}, A\&A, 565, A98

\bibitem[{{Larsen} {et~al}\mbox{.}(2014{\natexlab{b}}){Larsen}, {Brodie},
  {Grundahl}, \& {Strader}}]{larsen14}
{Larsen} S.~S., {Brodie} J.~P., {Grundahl} F., {Strader} J.,
  2014{\natexlab{b}}, ApJ, 797, 15

\bibitem[{{Larsen}, {Brodie} \& {Strader}(2012){Larsen}, {Brodie}, \&
  {Strader}}]{larsen12}
{Larsen} S.~S., {Brodie} J.~P., {Strader} J., 2012, A\&A, 546, A53

\bibitem[{{Mackey} {et~al}\mbox{.}(2008){Mackey}, {Broby Nielsen}, {Ferguson},
  \& {Richardson}}]{mackey08}
{Mackey} A.~D., {Broby Nielsen} P., {Ferguson} A.~M.~N., {Richardson} J.~C.,
  2008, ApJL, 681, L17

\bibitem[{{Marino} {et~al}\mbox{.}(2016){Marino}, {Milone}, {Casagrande},
  {Collet}, {Dotter}, {Johnson}, {Lind}, {Bedin}, {Jerjen}, {Aparicio}, \&
  {Sbordone}}]{marino16}
{Marino} A.~F. {et~al.}, 2016, MNRAS, 459, 610

\bibitem[{{Milone} {et~al}\mbox{.}(2012){Milone}, {Piotto}, {Bedin},
  {Aparicio}, {Anderson}, {Sarajedini}, {Marino}, {Moretti}, {Davies},
  {Chaboyer}, {Dotter}, {Hempel}, {Mar{\'{\i}}n-Franch}, {Majewski}, {Paust},
  {Reid}, {Rosenberg}, \& {Siegel}}]{milone12}
{Milone} A. {et~al.}, 2012, A\&A, 540, A16

\bibitem[{{Milone} {et~al}\mbox{.}(2009){Milone}, {Bedin}, {Piotto}, \&
  {Anderson}}]{milone09}
{Milone} A.~P., {Bedin} L.~R., {Piotto} G., {Anderson} J., 2009, A\&A, 497, 755

\bibitem[{{Milone} {et~al}\mbox{.}(2015){Milone}, {Bedin}, {Piotto}, {Marino},
  {Cassisi}, {Bellini}, {Jerjen}, {Pietrinferni}, {Aparicio}, \&
  {Rich}}]{milone15}
{Milone} A.~P. {et~al.}, 2015, MNRAS, 450, 3750

\bibitem[{{Monelli} {et~al}\mbox{.}(2013){Monelli}, {Milone}, {Stetson},
  {Marino}, {Cassisi}, {del Pino Molina}, {Salaris}, {Aparicio}, {Asplund},
  {Grundahl}, {Piotto}, {Weiss}, {Carrera}, {Cebri{\'a}n}, {Murabito},
  {Pietrinferni}, \& {Sbordone}}]{monelli13}
{Monelli} M. {et~al.}, 2013, MNRAS, 431, 2126

\bibitem[{{Mucciarelli} {et~al}\mbox{.}(2008){Mucciarelli}, {Carretta},
  {Origlia}, \& {Ferraro}}]{mucciarelli08}
{Mucciarelli} A., {Carretta} E., {Origlia} L., {Ferraro} F.~R., 2008, AJ, 136,
  375

\bibitem[{{Mucciarelli} {et~al}\mbox{.}(2011){Mucciarelli}, {Cristallo},
  {Brocato}, {Pasquini}, {Straniero}, {Caffau}, {Raimondo}, {Kaufer},
  {Musella}, {Ripepi}, {Romaniello}, \& {Walker}}]{mucciarelli11}
{Mucciarelli} A. {et~al.}, 2011, MNRAS, 413, 837

\bibitem[{{Mucciarelli} {et~al}\mbox{.}(2014){Mucciarelli}, {Dalessandro},
  {Ferraro}, {Origlia}, \& {Lanzoni}}]{mucciarelli14}
{Mucciarelli} A., {Dalessandro} E., {Ferraro} F.~R., {Origlia} L., {Lanzoni}
  B., 2014, ApJ, 793, L6

\bibitem[{{Mucciarelli} {et~al}\mbox{.}(2009){Mucciarelli}, {Origlia},
  {Ferraro}, \& {Pancino}}]{mucciarelli09}
{Mucciarelli} A., {Origlia} L., {Ferraro} F.~R., {Pancino} E., 2009, ApJ, 695,
  L134

\bibitem[{{Niederhofer} {et~al}\mbox{.}(2017{\natexlab{a}}){Niederhofer},
  {Bastian}, {Kozhurina-Platais}, {Larsen}, {Hollyhead}, {Lardo},
  {Cabrera-Ziri}, {Kacharov}, {Platais}, {Salaris}, {Cordero}, {Dalessandro},
  {Geisler}, {Hilker}, {Li}, {Mackey}, \& {Mucciarelli}}]{paperII}
{Niederhofer} F. {et~al.}, 2017{\natexlab{a}}, MNRAS, 465, 4159

\bibitem[{{Niederhofer} {et~al}\mbox{.}(2017{\natexlab{b}}){Niederhofer},
  {Bastian}, {Kozhurina-Platais}, {Larsen}, {Salaris}, {Dalessandro},
  {Mucciarelli}, {Cabrera-Ziri}, {Cordero}, {Geisler}, {Hilker}, {Hollyhead},
  {Kacharov}, {Lardo}, {Li}, {Mackey}, \& {Platais}}]{paperI}
{Niederhofer} F. {et~al.}, 2017{\natexlab{b}}, MNRAS, 464, 94

\bibitem[{{Niederhofer} {et~al}\mbox{.}(2015){Niederhofer}, {Georgy},
  {Bastian}, \& {Ekstr{\"o}m}}]{niederhofer15}
{Niederhofer} F., {Georgy} C., {Bastian} N., {Ekstr{\"o}m} S., 2015, MNRAS,
  453, 2070

\bibitem[{{Piotto} {et~al}\mbox{.}(2015){Piotto}, {Milone}, {Bedin},
  {Anderson}, {King}, {Marino}, {Nardiello}, {Aparicio}, {Barbuy}, {Bellini},
  {Brown}, {Cassisi}, {Cool}, {Cunial}, {Dalessandro}, {D'Antona}, {Ferraro},
  {Hidalgo}, {Lanzoni}, {Monelli}, {Ortolani}, {Renzini}, {Salaris},
  {Sarajedini}, {van der Marel}, {Vesperini}, \& {Zoccali}}]{piotto15}
{Piotto} G. {et~al.}, 2015, AJ, 149, 91

\bibitem[{{Rubele}, {Kerber} \& {Girardi}(2010){Rubele}, {Kerber}, \&
  {Girardi}}]{rubele10}
{Rubele} S., {Kerber} L., {Girardi} L., 2010, MNRAS, 403, 1156

\bibitem[{{Salinas} \& {Strader}(2015)}]{salinas15}
{Salinas} R., {Strader} J., 2015, ApJ, 809, 169

\bibitem[{{Sbordone} {et~al}\mbox{.}(2004){Sbordone}, {Bonifacio}, {Castelli},
  \& {Kurucz}}]{sbordone04}
{Sbordone} L., {Bonifacio} P., {Castelli} F., {Kurucz} R.~L., 2004, Memorie
  della Societa Astronomica Italiana Supplementi, 5, 93

\bibitem[{{Sbordone} {et~al}\mbox{.}(2011){Sbordone}, {Salaris}, {Weiss}, \&
  {Cassisi}}]{sbordone11}
{Sbordone} L., {Salaris} M., {Weiss} A., {Cassisi} S., 2011, A\&A, 534, A9

\bibitem[{{Stanimirovi{\'c}}, {Staveley-Smith} \&
  {Jones}(2004){Stanimirovi{\'c}}, {Staveley-Smith}, \&
  {Jones}}]{stanimirovic04}
{Stanimirovi{\'c}} S., {Staveley-Smith} L., {Jones} P.~A., 2004, ApJ, 604, 176

\bibitem[{{Villanova} {et~al}\mbox{.}(2013){Villanova}, {Geisler}, {Carraro},
  {Moni Bidin}, \& {Mu{\~n}oz}}]{villanova13}
{Villanova} S., {Geisler} D., {Carraro} G., {Moni Bidin} C., {Mu{\~n}oz} C.,
  2013, ApJ, 778, 186

\bibitem[{{Wu} {et~al}\mbox{.}(2016){Wu}, {Li}, {de Grijs}, \& {Deng}}]{wu16}
{Wu} X., {Li} C., {de Grijs} R., {Deng} L., 2016, ApJL, 826, L14

\bibitem[{{Yong}, {Grundahl} \& {Norris}(2015){Yong}, {Grundahl}, \&
  {Norris}}]{yong15}
{Yong} D., {Grundahl} F., {Norris} J.~E., 2015, MNRAS, 446, 3319

\end{thebibliography}

\label{lastpage}
\end{document}